\documentclass{appolb}
\usepackage{epsfig}
\usepackage{rotate}

\begin{document}
\renewcommand{\textfraction}{0.00000000001}
\renewcommand{\floatpagefraction}{1.0}
\title{Photoproduction of Mesons off Light Nuclei\\
 - the Search for $\eta$-Mesic Nuclei -
\thanks{International Symposium on Mesic Nuclei, Cracow June 2010}}
\author{B. Krusche, F. Pheron, and Y. Magrbhi
\address{Department of Physics, University of Basel,
Ch-4056 Basel, Switzerland}\\
for\\
the Crystal Ball/TAPS collaboration}
\date{\today}
\maketitle
\begin{abstract}
Photoproduction of $\eta$-mesons off light nuclei (d, $^3$He, $^7$Li) has 
been measured at the tagged photon beam of the Mainz MAMI accelerator with 
the combined Crystal Ball/TAPS detection system. Special attention was given
to the threshold behavior of the reactions in view of possible indications
for the formation of (quasi-)bound $\eta$-nucleus states, so-called 
$\eta$-mesic nuclei. A very strong threshold enhancement of coherent
$\eta$-photoproduction off $^3$He was found and coherent $\eta$-photoproduction
off $^7$Li was observed for the first time. Preliminary results will be
discussed.
\end{abstract}
  
\section{Introduction}
The interaction of mesons with nucleons and nuclei is a major source 
for our knowledge of the strong interaction. Elastic and inelastic
reactions using secondary pion and kaon beams have revealed many details
of the nucleon - meson potentials. However, secondary meson beams are only
available for long-lived, preferentially charged mesons. Much less is
known for short-lived mesons such as for example the $\eta$ and $\eta '$ 
mesons. Their interactions with nuclei can be studied only in indirect ways
when the mesons are first produced in a nucleus from the interaction of some
incident beam and then subsequently undergo final state interaction (FSI) 
in the same nucleus. 

A very interesting question is, whether the properties of the strong 
interaction allow the formation of meson - nucleus bound states, which
depend very sensitively on the strength of the interaction. Pionic atoms
are well established and deeply bound pionic states have recently been studied
by Geissel et al. \cite{Geissel_02,Geissel_02a}, yielding important results on
the in-medium properties of the pion decay constant $f^{\star}_{\pi}$
\cite{Kienle_04}, which are directly related to the predicted decrease of the 
chiral condensate in nuclear matter. In this case the superposition of the 
repulsive s-wave $\pi^-$ - nucleus interaction with the attractive Coulomb
interaction gives rise to bound states. 

\begin{figure}[th]
\begin{center}
\epsfig{figure=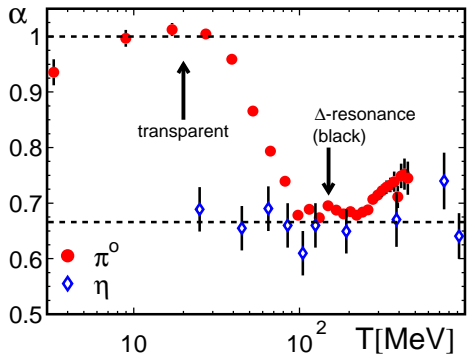,width=6cm}
\caption{Scaling parameter $\alpha$ as function of mesons kinetic energy
for $\pi^o$ and $\eta$ mesons.
\label{fig:scaling}
}
\end{center}
\end{figure}
Neutral mesons on the other hand can form bound states only in case of a 
sufficiently attractive strong meson - nucleus interaction.     
It is well known, that the pion - nucleus interaction for slow pions
is weak and can certainly not produce bound states. The typical energy 
dependence can be extracted for example from the strength of the final 
state interaction. This is shown in Fig.~\ref{fig:scaling} for the
absorption properties of photon induced production of $\pi^o$-mesons.
The production cross sections as function of the atomic mass number $A$
have been fitted with the simple Ansatz
\begin{equation}
\frac{d\sigma}{dT}(T)\propto A^{\alpha(T)}\;\;\; ,
\end{equation}
where $\alpha$ is the kinetic energy of the pions. A value of $\alpha$ 
close to unity corresponds to a cross section scaling with the volume of 
the nucleus, i.e. with vanishing absorption (very weak FSI), while a value 
$\approx$2/3 indicates surface proportionality corresponding to strong 
absorption. As expected, pions show strong FSI at kinetic energies large 
enough to excite the $\Delta$(1232) nucleon resonance, but are almost
undisturbed for energies below the $\Delta$ excitation threshold.
The situation is completely different for $\eta$-mesons, which show strong
absorption for all kinetic energies measured so far. The reason is the
overlap of the $s$-wave resonance S$_{11}$(1535), which couples strongly
to the $N\eta$-channel, with the production threshold. Eta-production
in the threshold region is therefore completely dominated by this resonance
\cite{Krusche_97}. 

Already in 1985 Bhalerao and Liu \cite{Bhalerao_85} performed coupled channel
calculations for the $\pi N\rightarrow\pi N$, $\pi N\rightarrow\eta N$, and
$\eta N\rightarrow\eta N$ reactions and found an attractive s-wave 
$\eta$N-interaction. Shortly later, Liu and Haider \cite{Liu_86} pointed out,
that for nuclei with A $>$ 10 this interaction may lead to the formation of
strongly bound $\eta$-nucleus systems which they termed {\it $\eta$-mesic}
nuclei. Experimental evidence for `heavy' $\eta$-mesic nuclei was e.g.
searched for in $A(\pi^+ ,p)\eta (A-1)$ reactions \cite{Chrien_88}, where
it should manifest itself by a kinematic peak from the two-body
$A(\pi^+ ,p)(A-1)_{\eta}$ process and in double pionic charge exchange
reactions \cite{Johnson_93}. However no conclusive evidence for the
existence of $\eta$-mesic nuclei was reported from such experiments.
A.I.Lebedev and V.A.Tryasuchev \cite{Lebedev_91} have pointed out that 
photon induced reactions are advantageous for the search of $\eta$-mesic
nuclei because they avoid initial state interaction effects. In contrast to 
hadronic induced reactions the photon can produce an $\eta$-meson with 
{\it any} of the nucleons.
Recently, Sokol et al. \cite{Sokol_01} claimed evidence for the formation of 
$\eta$-mesic nuclei with mass number $A=11$ (carbon, beryllium) in the 
$\gamma +^{12}$C reaction via 
\begin{equation}
\label{eq:leb}
\gamma + A
\rightarrow
N_{1} + (A-1)_{\eta}
\rightarrow
N_{1} + (N_{2}+\pi)+(A-2)
\end{equation}
where the $\eta$-meson is produced on the nucleon $N_{1}$, captured in the
rest nucleus $(A-1)$, which subsequently decays by emission of a nucleon-pion
pair.

During the last two decades the data basis for the $\eta$N-interaction was 
supplemented by new precise data in particular for $\eta$-photoproduction 
from the proton 
\cite{Krusche_95}-\cite{Prakhov_10},
the deuteron 
\cite{Krusche_95a}-\cite{Jaegle_08},
and He-nuclei 
\cite{Hejny_99}-\cite{Pfeiffer_04}.
This has prompted several groups to carry out new analyses of the 
$\eta$N-interaction. Most of them find a real part of the $\eta$N-scattering
length $a$, which is considerably larger than the original value from the 
work of Bhalerao and Liu \cite{Bhalerao_85} (a=0.27+$i$0.22). The more 
recent results for the real part of the scattering length span the entire 
range from 0.2 - 1. and most cluster between 0.5 - 0.8 (see e.g. 
\cite{Sibirtsev_02} and ref. therein). These larger values  
have prompted speculations about the existence of much lighter $\eta$-mesic 
nuclei than originally considered by Liu and Haider \cite{Liu_86}.
Light quasi-bound $\eta$-nucleus states have been sought in experiments 
investigating the threshold behavior of hadron induced $\eta$-production 
reactions. The idea is that quasi-bound states in the vicinity of the 
production threshold will give rise to an enhancement of the cross section 
relative to the expectation for phase space behavior. 
The threshold behavior of hadron induced $\eta$-production reactions has
been studied in detail in particular $pp\rightarrow pp\eta$ 
\cite{Calen_96}-\cite{Moskal_04}, 
$np\rightarrow d\eta$  \cite{Plouin_90,Calen_98}, 
$pd\rightarrow\eta  ^3\mbox{He}$ \cite{Mayer_96}, 
$dp\rightarrow\eta  ^3\mbox{He}$ \cite{Smyrski_07,Mersmann_07}, 
$\vec{d}d\rightarrow \eta ^4\mbox{He}$ \cite{Willis_97}, 
and $pd\rightarrow pd\eta$ \cite{Hibou_00}. All reactions show more or less 
pronounced threshold enhancements. 

\newpage
If quasi-bound states do exist they should show up as threshold enhancements 
independently of the initial state of the reaction. Photoproduction of 
$\eta$-mesons offers the advantage of a very well understood elementary 
reaction and rather clean experimental signals. In a previous experiment
Pfeiffer et al. \cite{Pfeiffer_04} searched for signatures of an $\eta$-mesic
state for the $^3$He system. Two different approaches were followed. 
Once a (quasi-)bound state has been formed
it may decay above threshold via emission of the $\eta$-meson, giving rise to
the threshold enhancement of coherent $\eta$-production. The $\eta$-meson
may also be re-captured by a nucleon, exciting it to the S$_{11}$-state which
decays with a $\approx$50~\% branching ratio to a nucleon - pion pair.
This decay channel should result in a peak-like structure in the excitation
function of nucleon - pion pairs, emitted back-to-back in the photon - $^3$He
center-of-momentum (cm) system. Some indication for both effects has been
reported by Pfeiffer et al. \cite{Pfeiffer_04}, however at the limit of
statistical significance.

The aim of the present experiments was therefore a large improvement of the
statistical quality of the previous experiment, and in addition the study 
of it in a different nuclear system namely $^7$Li. The selection criterion
for the nucleus is simple, due to the dominance of an iso-vector, spin-flip
amplitude in the S$_{11}$ excitation \cite{Krusche_03}, and thus in threshold
$\eta$-photoproduction, only nuclei with isospin $I\neq$0 and spin $J\neq$0
can have non-negligible reaction rates for the coherent 
$\gamma A\rightarrow A\eta$ process.

\section{Experimental setup}
\begin{figure}[th]
\centerline{\epsfig{figure=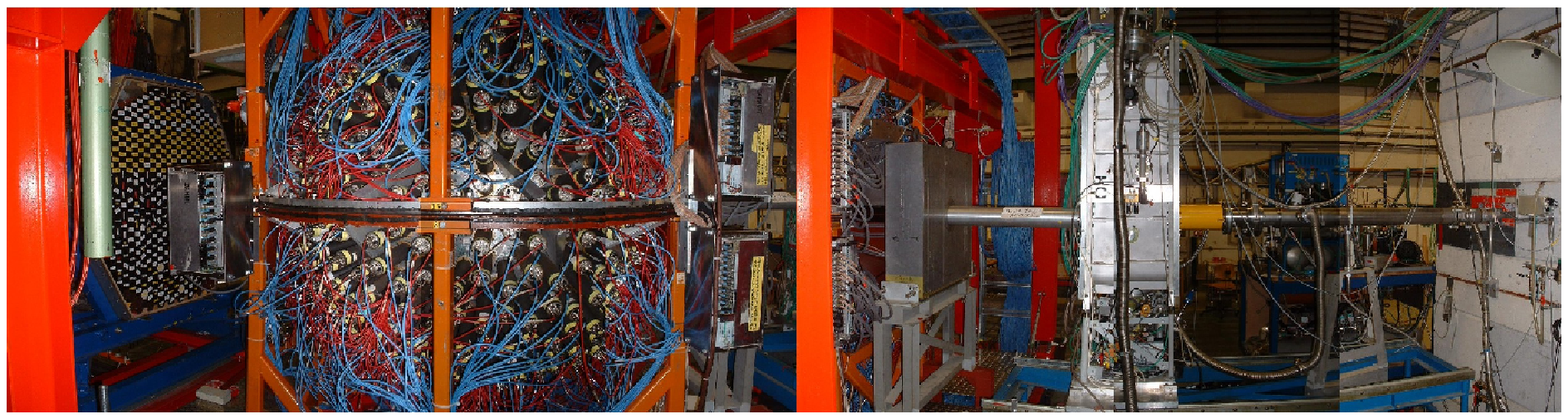,width=12.5cm}}
\vspace*{0.5cm}
\begin{center}
\epsfig{figure=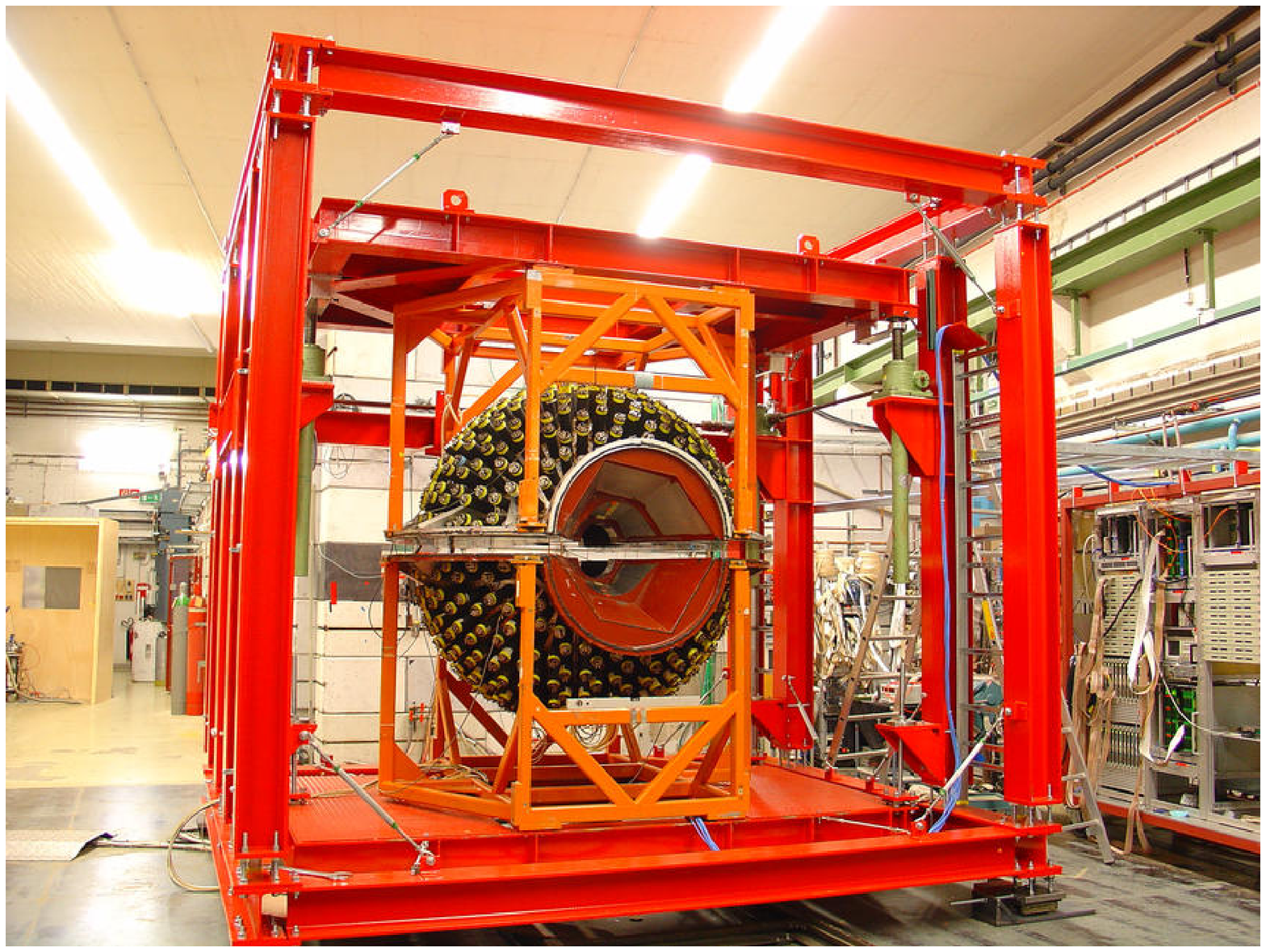,width=6.15cm}
\epsfig{figure=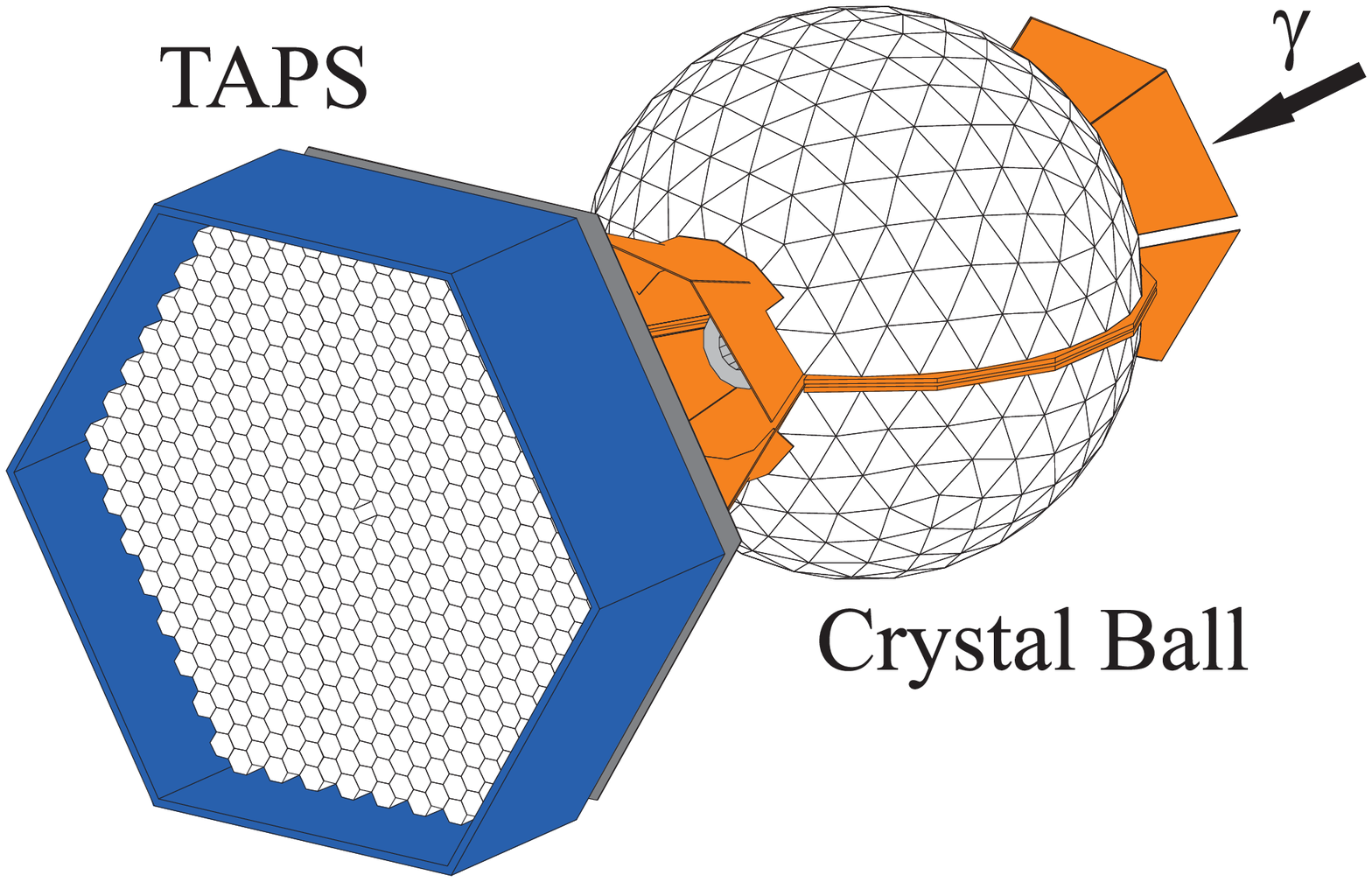,width=6.15cm}
\caption{Overview of the experimental setup. Upper part: beam coming from the
right, Crystal Ball at the left of the picture, TAPS forward wall at the very
left. 
Bottom part, left hand side: Crystal Ball in setup phase, right hand side:
schematic drawing of the calorimeter. 
\label{fig:setup}
}
\end{center}
\end{figure}

The experiments were done at the tagged photon beam of the Mainz MAMI
accelerator \cite{Herminghaus_83,Kaiser_08}. The electron beam of 1508 MeV 
energy was impinging on a copper radiator of 10$\mu$m thickness to produce 
bremsstrahl photons which were tagged with the Glasgow magnetic spectrometer 
\cite{Anthony_91} for photon energies between 0.45 GeV - 1.4 GeV. 
The target was a capton cylinder of 4.3 cm diameter and 5.3 cm length
filled with liquid $^3$He at a temperature of 2.6 K. The target density
was 0.07 nuclei/barn. The $\eta$-mesons were identified via their
$\eta\rightarrow 2\gamma$ and $\eta\rightarrow 3\pi^o\rightarrow 6\gamma$
decays. The decay photons were detected with an electromagnetic
calorimeter combining the Crystal Ball \cite{Oreglia_82} made of 672 NaI 
crystals with 384 BaF$_2$ crystals of the TAPS detector 
\cite{Novotny_91,Gabler_94} configured as a forward wall. The Ball was equipped
with an additional inner detector (plastic scintillators) for charged particle 
identification via the $\Delta E - E$ method and all modules of the TAPS
detector had individual plastic scintillators in front for the same purpose.
Pictures and a schematic overview of the setup are shown in Fig.
\ref{fig:setup}. The Crystal Ball covered the full azimuthal range for polar 
angles from 20$^o$ to 160$^o$, corresponding to 93~\% of the full solid angle.
The TAPS detector, mounted 1.07 m downstream from the target, covered 
polar angles from $\approx$2$^o$ to 21$^o$.  

\section{Data analysis}
Photoproduction of $\eta$-mesons was analyzed via their 2$\gamma$- and 
6$\gamma$-decay channels. For the analysis of the coherent process 
$\gamma^3\mbox{He}\rightarrow\eta^3\mbox{He}$ only events with exactly two
(respectively six) photons and no further hit in the detector were accepted.
This reduces the background from breakup reactions since events were the
participant nucleon is detected are suppressed. The $\eta$-mesons were
identified with a standard invariant mass analysis for the two photon decay.
Residual background under the $\eta$-peak was subtracted by a fit of the
simulated line shape and a background polynomial. In case of the 
$\eta\rightarrow 3\pi^o\rightarrow 6\gamma$ decay, the six photons were first
combined via a $\chi^2$-test to three pairs which are the best solution for
three $\pi^o$ invariant masses. A cut between 110 MeV - 150 MeV was made on 
these invariant masses. 
Subsequently the six photon invariant mass was
constructed, which in this case is an almost background free $\eta$ invariant
mass peak (the cross section for triple $\pi^o$ production not related to
$\eta$-decays is very small in the interesting range of incident photon
energies).

\begin{figure}[th]
\begin{center}
\epsfig{figure=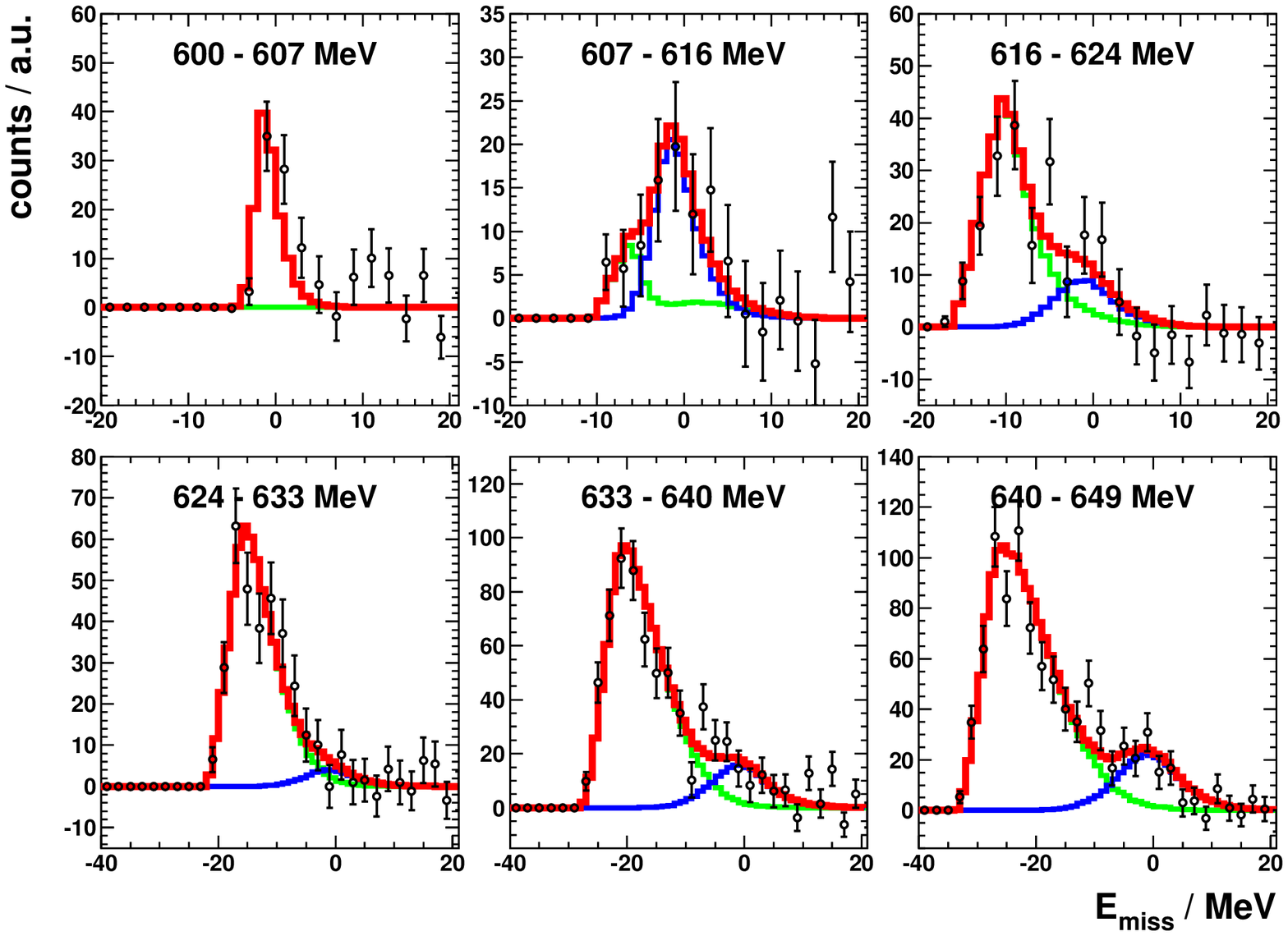,width=9cm}
\epsfig{figure=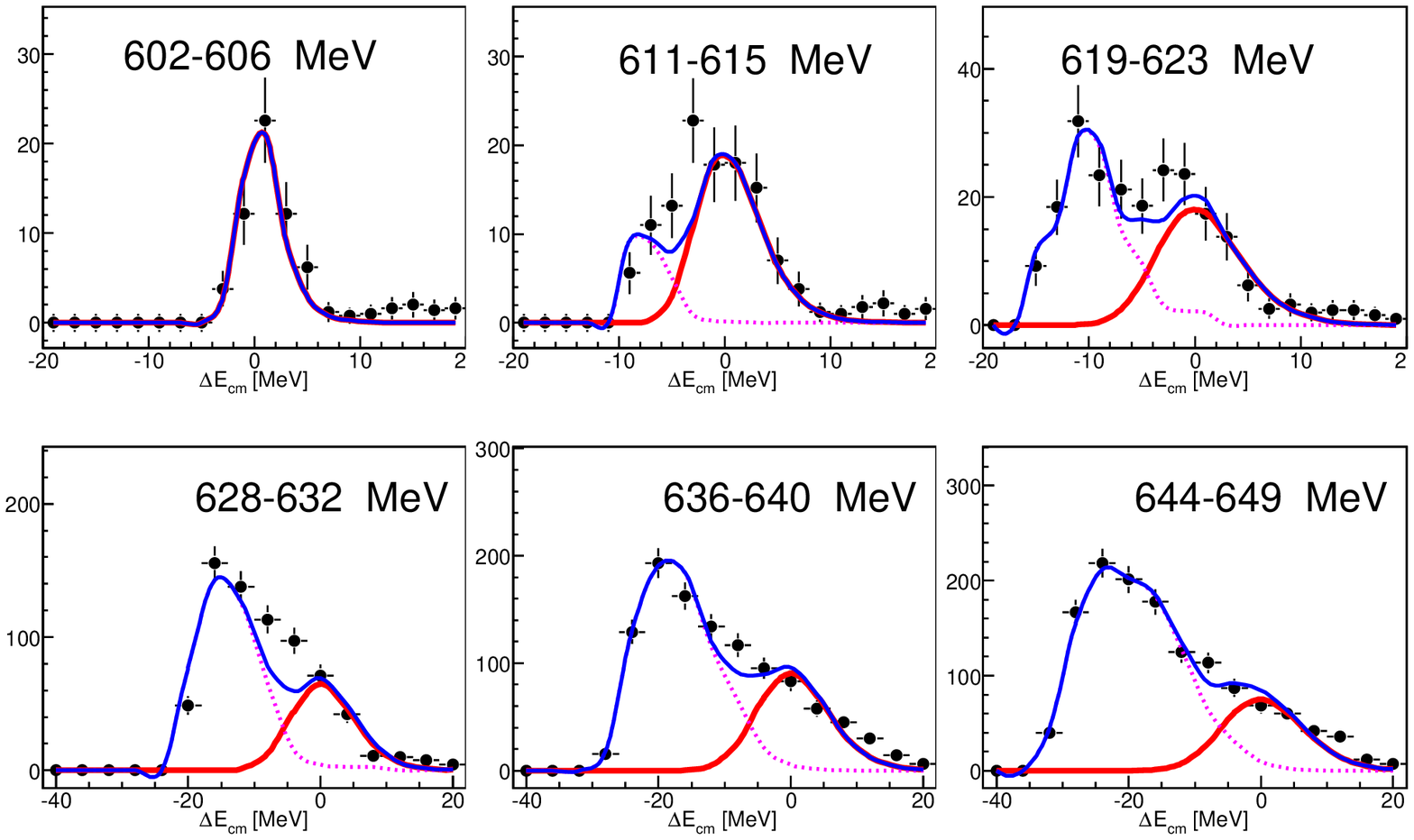,width=9cm}
\caption{Missing energy spectra. Upper part: results from the previous
experiment \cite{Pfeiffer_04}, bottom part: present experiment. In both cases
for the $\eta\rightarrow 2\gamma$ decay. Lines: simulated line shapes
of coherent signal (centered around $\Delta E$=0), breakup background (at
negative missing energies) and sum of both. 
\label{fig:miss}
}
\end{center}
\end{figure}

The next important step is then the separation of the coherent events from
residual background from breakup reactions were a nucleon is removed from
the Helium nucleus but not detected. This was done using
a missing energy analysis which compares the kinetic cm energies of the 
$\eta$-mesons extracted directly from the measured decay photons with the ones
calculated from the incident photon energy under the assumption of coherent
kinematics. The result is a peak around zero for coherent events and some
background structure at negative missing energies for breakup events.
The shape of both contributions has been generated with Monte Carlo simulations
using the GEANT4 package \cite{Geant_03}. This analysis was done independently for
each bin of incident photon energy and cm polar angle of the $\eta$-mesons.
The result of this analysis is compared in Fig.~\ref{fig:miss} to the results
of the previous experiment by Pfeiffer et al. \cite{Pfeiffer_04}. Due to the large
angle coverage of the present experiment not only the statistical quality of the
coherent signal is much improved. In addition, the relative contribution from
background events from breakup reactions is much reduced due to the detection
of the recoil nucleons. As in the previous experiment, the first energy bin
between coherent and breakup threshold shows a clean single of coherent
production, in good agreement with the simulated line shape. An identical
analysis was done for the six-photon channel and for the $^7$Li nucleus.

\section{Results}
\begin{figure}[th]
\begin{center}
\epsfig{figure=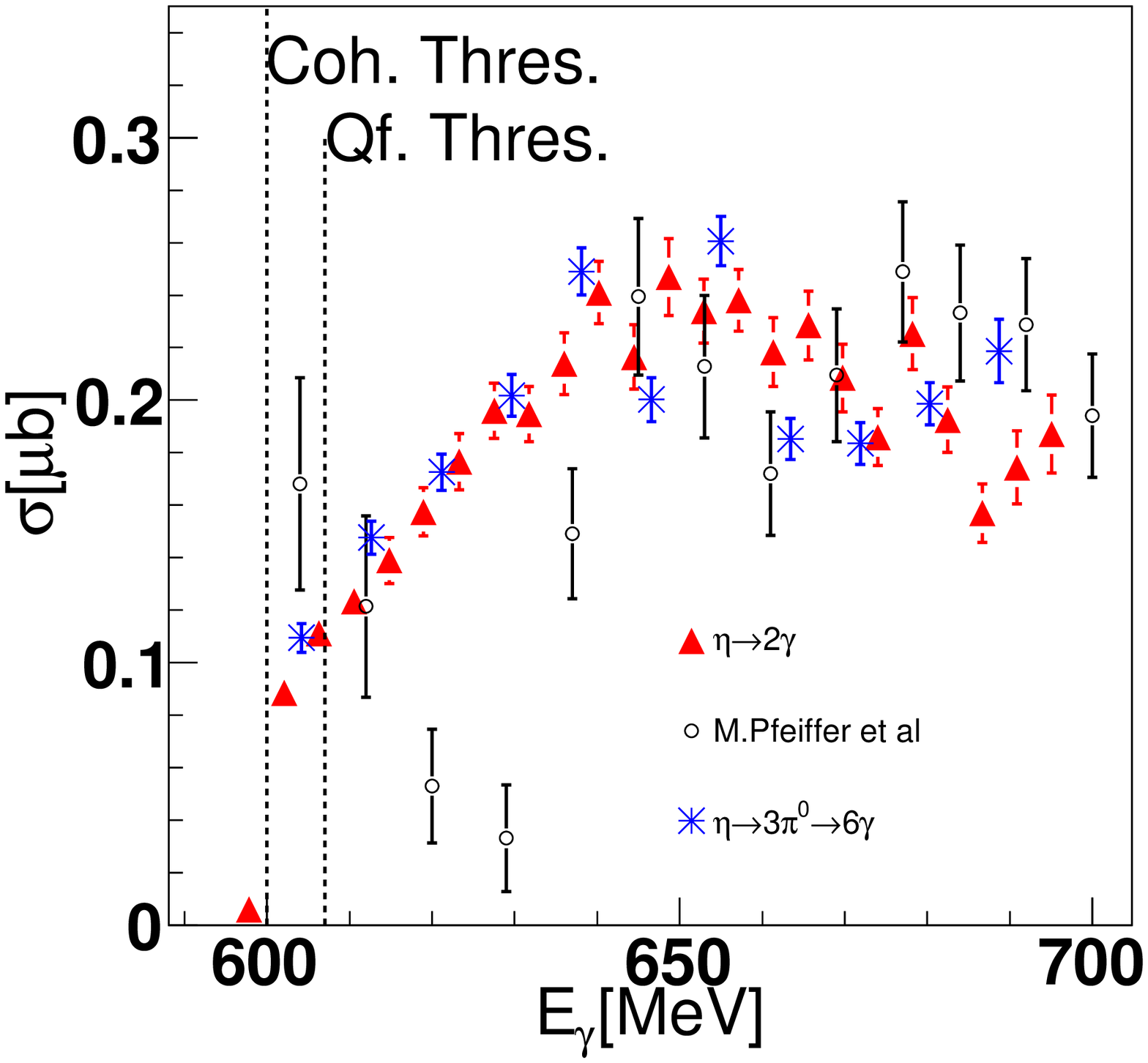,width=6.2cm}
\epsfig{figure=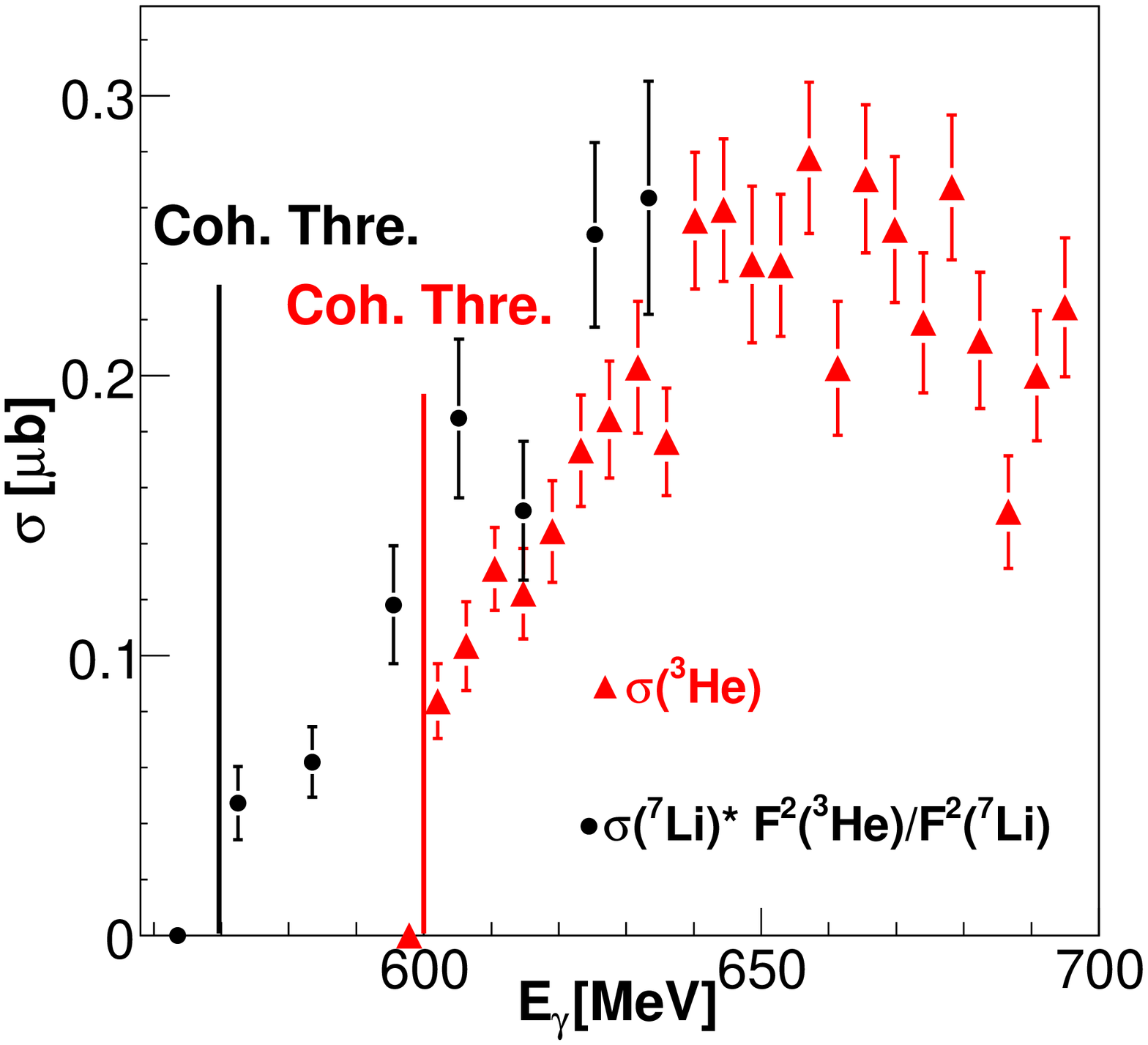,width=6.2cm}
\caption{Left hand side: total cross section for coherent $\eta$ production off
$^3$He from the two $\eta$-decay channels compared to the previous result from
Pfeiffer et al. \cite{Pfeiffer_04}. Right hand side: comparison of  
$\gamma^3\mbox{He}\rightarrow\eta^3\mbox{He}$ to 
$\gamma^7\mbox{Li}\rightarrow\eta^7\mbox{Li}$ scaled by the ratio of the
nuclear form factors. All results preliminary.
\label{fig:total}
}
\end{center}
\end{figure}

Preliminary results for the extracted total cross sections are summarized
in Fig.~\ref{fig:total}. The results from the two- and six-photon decays of the
$\eta$-mesons are in quite good agreement. Both excitation functions show an
extremely steep rise at coherent threshold, which cannot be resolved with the
incident photon energy resolution of 4 MeV defined by the width of the tagger
focal plane detectors. This result is very similar to the observation in the
hadron induced reaction $pd\rightarrow\eta^3\mbox{He}$ studied by Mersmann et al.
\cite{Mersmann_07}.
The independence of this effect on the initial state clearly
demonstrates 
its origin from final state properties and is thus an indication towards the 
existence of a (quasi-)bound state. The data are in reasonable agreement with 
the previous results \cite{Pfeiffer_04} except in the energy region 
from 620 - 630 MeV. Inspection of the missing energy spectra shows, that the
previous experiment within its low statistics and the presence of the large
background structure was consistent with a null result, while the present data
show a clear and statistically significant coherent contribution in this range.

The right hand side of the figure shows a comparison of the total cross sections
for coherent production off $^3$He and $^7$Li. The latter is scaled up by
the ratio of the form factors of the nuclei which differs by roughly one order 
of magnitude so that the Li cross sections are on the order of only a few ten nb.
This is the first positive signal for coherent $\eta$ photoproduction for
any nucleus with mass beyond $A$=3. The fact that after correction for the nuclear
form factor the cross sections for the two nuclei are very similar is not
surprising. Coherent production for spin-/isospin independent reactions like
$\pi^o$-photoproduction in the $\Delta$ range scales with the atomic mass
numbers. However, since in
$\eta$-photoproduction the isovector, spin-flip component is dominant, for both
nuclei the largest contributions comes from the single unpaired nucleon.
This is the $s_{1/2}$ neutron for $^3$He and the $p_{3/2}$ proton for $^7$Li.
Also the Li-data show a steep rise at coherent threshold, however, less pronounced
as in the Helium case.

\vspace*{0.5cm}
\begin{figure}[th]
\begin{center}
\epsfig{figure=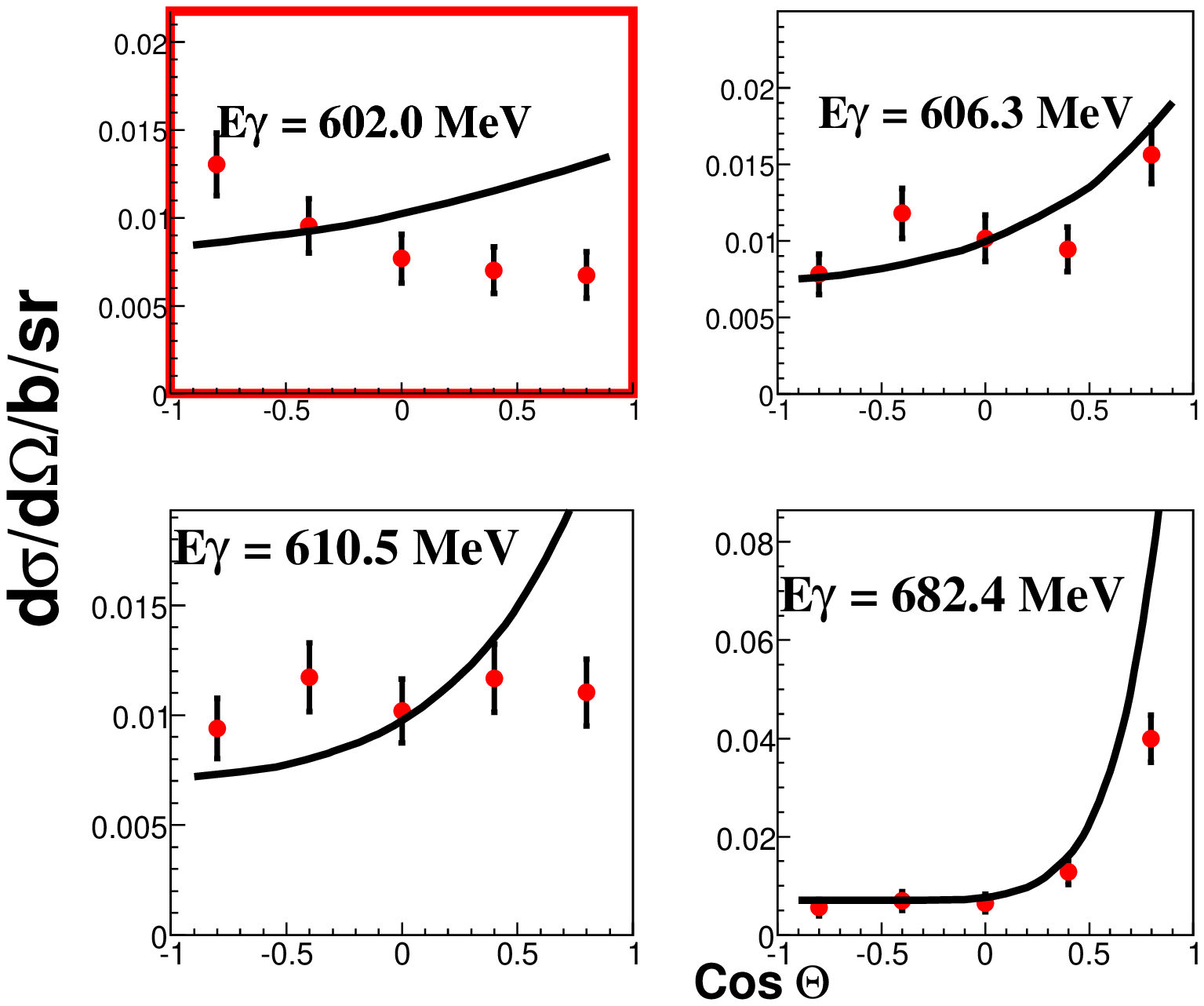,width=5.5cm}\hspace*{1cm}
\epsfig{figure=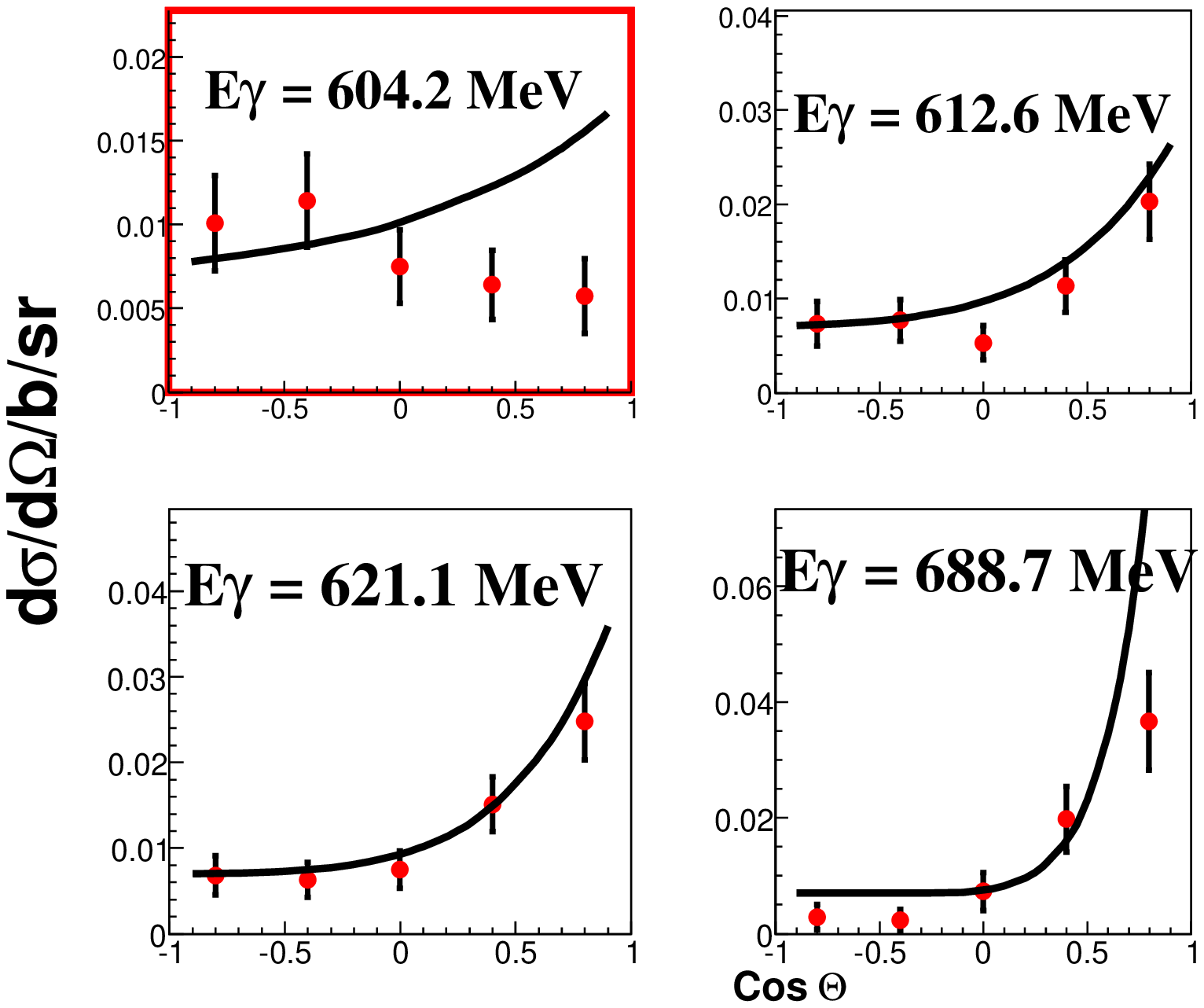,width=5.5cm}
\caption{Angular distributions for $\gamma^3\mbox{He}\rightarrow\eta^3\mbox{He}$.
Left hand side: two-photon channel, right hand side: six-photon decay. The curves
show the behavior expected from the nuclear form factor. All results preliminary.
\label{fig:diff}
}
\end{center}
\end{figure}
 
Angular distributions for the coherent reaction off $^3$He are summarized in 
Fig.~\ref{fig:diff}. Since the elementary reactions of $\eta$-photoproduction off the
nucleon are not much angular dependent in the energy range of interest one expects
that the main angular dependence arises from the nuclear form factor $F^2(q^2)$.
As demonstrated in the figure, this is the case except for the immediate vicinity
of the threshold. This is a further indication that FSI effects are dominated
in this region.  

\begin{figure}[th]
\begin{center}
\epsfig{figure=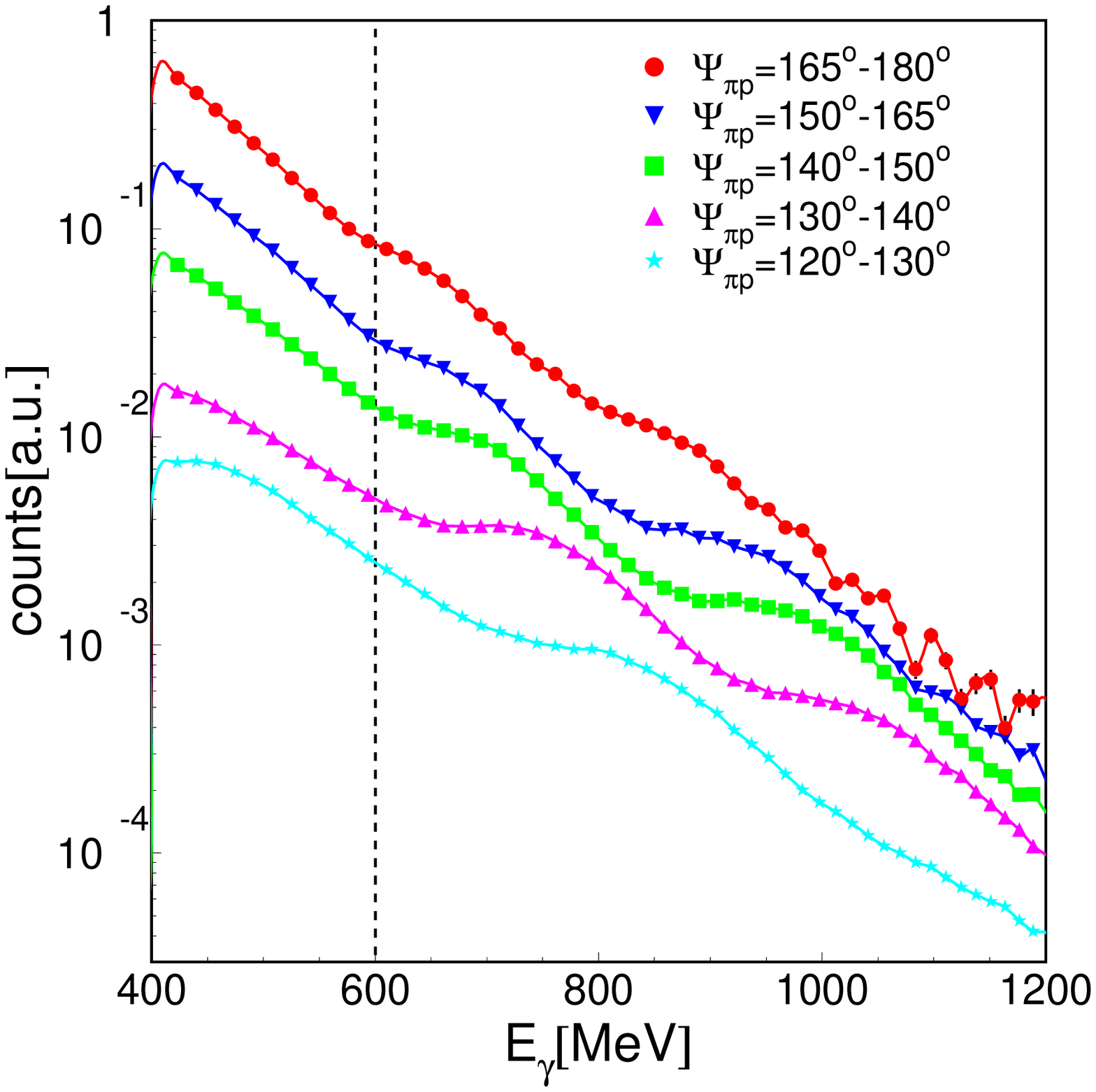,width=5.5cm}\hspace*{1cm}
\epsfig{figure=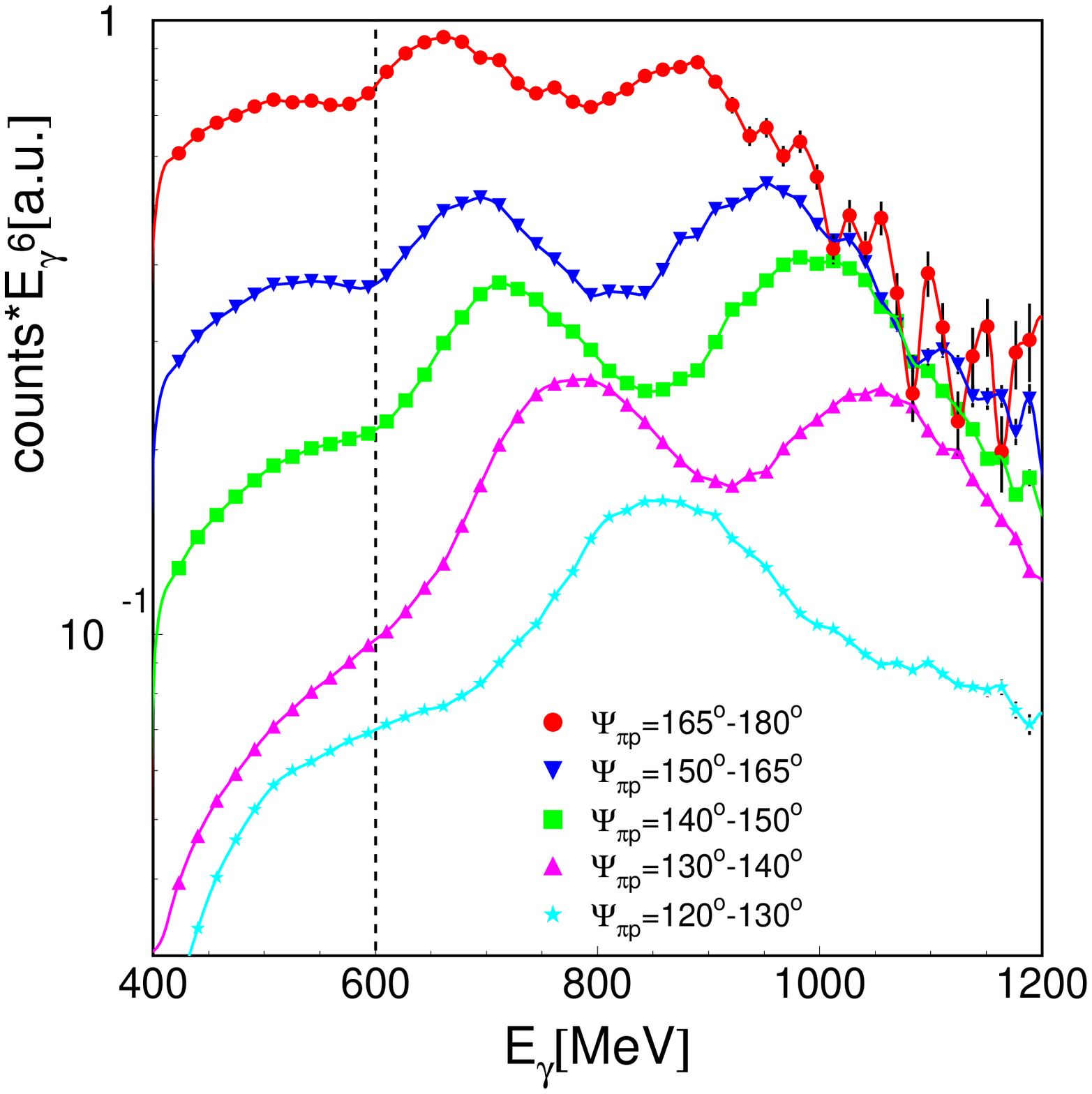,width=5.5cm}
\caption{Excitation functions of $\pi^o$ - proton pairs with different opening
angles $\psi$ in the photon - $^3$He cm frame. Left hand side: excitation
functions, right hand side: multiplied with $E_{\gamma}^6$ to remove the overall
energy dependence.
\label{fig:pi0}
}
\end{center}
\end{figure}
A further signal for the formation of a (quasi-)bound state discussed 
by Pfeiffer et al. \cite{Pfeiffer_04} was a narrow peak around the
$\eta$-production threshold in the excitation function of $\pi^o$ - proton
back-to-back pairs in the photon - $^3$He cm system. The kinematics for the
decay of a (quasi-)bound state via re-capture of the $\eta$ by a nucleon and
subsequent decay of the S$_{11}$(1535) into a pion -nucleon pair has been
simulated. Events from the data have been selected in a way to maximize the ratio
of this signal with respect to quasi-free $\pi^o$ production. The excitation
functions are shown in Fig.~\ref{fig:pi0}. Following the same procedure as 
in \cite{Pfeiffer_04} by subtracting properly scaled excitation functions for 
opening angles between 150$^o$ - 165$^o$ from the data for 165$^o$ - 180$^o$
indeed reproduces the narrow structure discussed in \cite{Pfeiffer_04}. 
However, the new data of much better statistical quality covering a larger 
incident photon energy range clearly show the appearance of two structures in 
this excitation functions arising from the third and second resonance region in 
quasi-free $\pi^o$ photoproduction. These structures move in
incident photon energy as function of the selected opening angle between
the $\pi^o$ and the proton, which is a trivial kinematic effect. Unfortunately,
the subtraction of the excitation functions of the expected signal bin and the 
background bin accidentally generates an artificial structure right at the 
$\eta$-threshold.

\section{Conclusions}
Coherent photoproduction of $\eta$-mesons has been studied in the threshold
region for $^3$He and $^7$Li nuclei. The results for $^3$He are much improved in
terms of statistical quality with respect to a previous measurement
\cite{Pfeiffer_04} and confirm the extremely steep, almost step-like rise of the
excitation function at the coherent threshold which is evidence for strong
final state interaction. Together with the results from the hadron induced
reaction $pd\rightarrow\eta^3\mbox{He}$ \cite{Mersmann_07}, which show a similar
behavior, this supports the formation of a (quasi-)bound $\eta$ - nucleus
state. The angular distributions also show the previously observed deviation from
the expected form factor dependence in the threshold region. A similar measurement
for $^7$Li produced the first ever data for coherent $\eta$-photoproduction off
a nucleus beyond the $A$=3 mass range and show also a strong threshold enhancement
also less pronounced than for $^3$He. The previously observed \cite{Pfeiffer_04}
threshold structure in the $\pi^o$ - p back-to-back emission excitation function
exists, but is most likely an artefact from quasi-free $\pi^o$ production.

\begin{center}
{\bf Acknowledgments}
\end{center}

\noindent{This} work was supported by Schweizerischer Nationalfonds and 
Deutsche Forschungsgemeinschaft (SFB 443).

\end{document}